\documentclass{article}
\usepackage{spconf,amsmath,graphicx}
\usepackage{enumitem}

\title{A joint detection-classification model for audio tagging of weakly labelled data}
\name{Qiuqiang Kong, Yong Xu, Wenwu Wang, Mark D. Plumbley\thanks{This research is supported by EPSRC grant EP/N014111/1 ``Making Sense of Sounds'' and research scholarship from the China Scholarship Council (CSC).}}
\address{ Center for Vision, Speech and Signal Processing (CVSSP) \\  University of Surrey\\ \{q.kong, yong.xu, w.wang, m.plumbley\}@surrey.ac.uk }

\begin{document}
%
\maketitle
\begin{abstract}
Audio tagging aims to assign one or several tags to an audio clip. Most of the datasets are \textit{weakly labelled}, which means only the tags of the clip are known, without knowing the occurrence time of the tags. The labeling of an audio clip is often based on the audio events in the clip and no event level label is provided to the user. Previous works have used the \textit{bag of frames} model assume the tags occur all the time, which is not the case in practice. We propose a \textit{joint detection-classification} (JDC) model to detect and classify the audio clip simultaneously. The JDC model has the ability to attend to informative and ignore uninformative sounds. Then only informative regions are used for classification. Experimental results on the ``CHiME Home'' dataset show that the JDC model reduces the equal error rate (EER) from 19.0\% to 16.9\%. More interestingly, the audio event detector is trained successfully without needing the event level label.

\end{abstract}
\section{Introduction}
\label{sec:intro}

Audio tagging aims to assign an audio clip with one or several tags. The clips are typically short segments such as 10 seconds of a long recording. Audio tagging has applications such as audio retrieval \cite{guo2003content} and audio classification \cite{stowell2015detection}. Accurate audio tagging relies on the amount of labelled audio data, including \textit{clip level} labelled audio data and \textit{event level} labelled audio data. In clip level labelling, each audio clip is labelled with one or several tags without indicating their occurrence time, while in event level labelling, each audio clip is labelled with both tags and their occurrence time. We refer to the clip level labelled data as \textit{weakly labelled data} \cite{mann2010generalized}.  Figure 1 shows an audio clip and its event level label. The clip\footnote{The clip is CR\_lounge\_220110\_0731.s0\_chunk39 from the ``CHiME Home'' dataset.} level label is ``children speech'', ``percussion'' and ``other sounds''.

Event level labelled data is scarce and insufficient for industrial use \cite{kumar2016audio}. On the other hand, the amount of weakly labelled data on the internet is exploding in recent years. Every minute, over 300 hours videos are uploaded to YouTube\footnote{http://tubularinsights.com/youtube-300-hours/}. Each of the clip has a few tags roughly tagged by the uploaders describing the content such as ``music", ``dogs", ``Obama". These weakly labelled data have huge potential values but are currently difficult to use. First, the tags only indicate the existence but not the occurrence times of the events. Second, the duration of audio events varies from events to events. Some events such as ``music'' may last for a few minutes, while other events such as ``gun shot'' may only last for hundreds of milliseconds. Third, overlapping events are difficult to classify. These difficulties limit the use of weakly labelled data. 

\begin{figure}[t]
  \centering
  \includegraphics[width=\columnwidth]{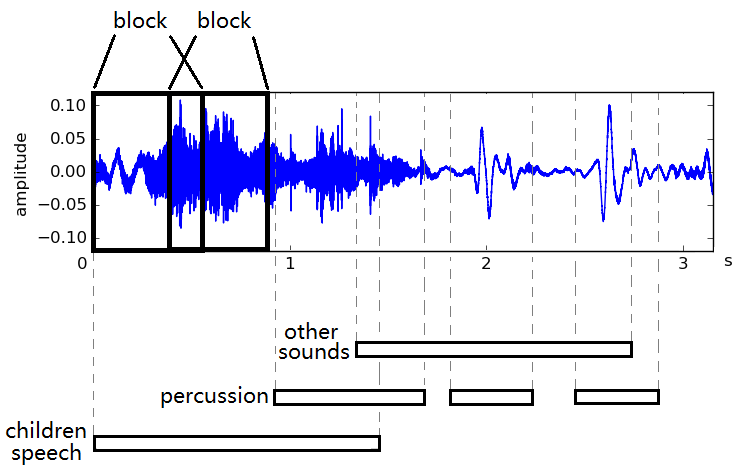}
  \caption[Caption for LOF]{Waveform of an audio clip \ and its event level label. Clip level label: ``children speech'', ``percussion'', ``other sounds''. }
  \label{fig:waveform}
\end{figure}

Previous work in audio tagging includes the \textit{bag of blocks} (BOB) model \cite{stowell2015detection}\cite{mesaros2016tut} and the \textit{global model} \cite{choi2016automatic}. The BOB model is a generalization of the \textit{bag of frames model} \cite{stowell2015detection}. In the BOB model, each audio clip is divided into overlapping blocks (Figure 1). Each block contains one or several frames. In the training phase, for a machine learning system, each block is assigned with all tags in the clip as ground truth for training. In the recognition phase, the predicted tags from all blocks in a clip are used to estimate a predicted probability of the tag in that clip. The BOB model is based on an assumption that tags occur in all the blocks, which is however not the case in practice. 

In contrast with the bag of blocks model, the global models feeds the whole clip into the model, without dividing the clip into blocks (Figure 2(b)). The advantage of the global model over the BOB model is that there is no ``tags occur all the time'' assumption. In addition, long-term context is considered. Inspired by image classification \cite{krizhevsky2012imagenet}, the global model can be for example, a convolutional neural network (CNN) applied on the clip's spectrogram \cite{choi2016automatic}. 

However, neither the BOB model nor the global model looks for the audio events as humans do before tagging an audio clip. In this paper, we propose a \textit{joint detection-classification} (JDC) model inspired by humans perception \cite{bregman1994auditory}. Humans use two steps to label an audio, a detection step and a classification step \cite{bregman1994auditory}. In the detection step, humans listen when to \textit{attend} to sounds and when to \textit{ignore} sounds. They attend to informative audio events and ignore uninformative sounds or silences. In a classification task, humans only tag informative events. Our JDC model is trained on weakly labelled data, without needing event level labelled data. 

Our work is organized as follows, Section 2 introduces related works. Section 3 describes the JDC model. Section 4 shows experimental results. Section 5 gives conclusions.

\section{Related Work}
\label{sec:Related Work}

Audio classification and detection have obtained increasing attention in recent years, such as the open challenges for audio classification, audio event detection and audio tagging including CLEAR 2007 \cite{ooi2007evaluation}, DCASE2013 \cite{stowell2015detection} and DCASE2016 \cite{mesaros2016tut}. 

For audio classification and audio tagging, Mel frequency cepstrum coefficient (MFCC) and Guassian mixture model (GMM) is widely used as baseline \cite{mesaros2016tut}. Recent methods include deep neural networks (DNNs) and convolutional neural networks (CNNs), with inputs varying from Mel energy \cite{valentidcase}, spectrogram \cite{han2016acoustic} to constant Q transoform (CQT) \cite{lidy2016cqt}. For non neural network models, nonnegative matrix factorization (NMF) on the spectrogram \cite{bisot2016acoustic} was used and obtained competitive result in the DCASE2016 Challenge. Methods combining the BOB model and the global global are proposed in \cite{kumar2016audio}. 

For audio event detection, generative models such as NMF \cite{dessein2013real}, probabilistic latent component analysis (PLCA) \cite{benetos2016detection} and hidden Markov model (HMM) \cite{heittola2013context} are used with inspiration from music transcription. Neural network methods including recurrent neural networks (RNNs) \cite{leglaive2015singing} and bidictionary long-short term memory (BLSTM) \cite{parascandolo2016recurrent} are also used for event detection. 

\section{{Joint detection$-$classification model}}
\label{sec:Joint detection classification model}

For the audio tagging task, assume there are $ K $ different audio tags. We denote the labelled target of an audio clip as $ t \in \{0,1\}^{K} $ where $ t_{k} \in \{0,1\} $ represents the existence of the $ k $-th tag. For the bag of blocks model (Figure 2(a)), the audio clip is divided into overlapped blocks $ x_{m}, m=1,2,...,M $. We denote $ y_{mk} $ as the predicted occurrence probability of the $ k $-th tag on the $ m $-th block. In the training phase, the ``tags occur in all blocks'' assumption is applied and each block is assigned with the clip's tags as ground truth $ t_{mk}=t_{k} $. The model is trained by minimizing the loss function

\begin{equation} \label{eqBobLoss}
loss = \frac{1}{M} \sum_{k=1}^{K} \sum_{m=1}^{M} d(y_{km},t_{mk})
\end{equation}

\noindent
where $ d(\cdot,\cdot) $ is the binary cross-entropy error. 

In the recognition phase, we obtain $ y_{km} $ on each $ k $-th tag and each $ m $-th block by feeding the $ m $-th block to the model. The occurrence probability of the $ k $-th tag in the clip $ p_{k} $ is obtained by averaging $ y_{km} $ over the blocks 

\begin{equation} \label{eqBobPred}
p_{k} = \frac{1}{M} \sum_{m=1}^{M} y_{km}
\end{equation}

\begin{figure*}[t]
  \centering
  \centerline{\includegraphics[width=\textwidth]{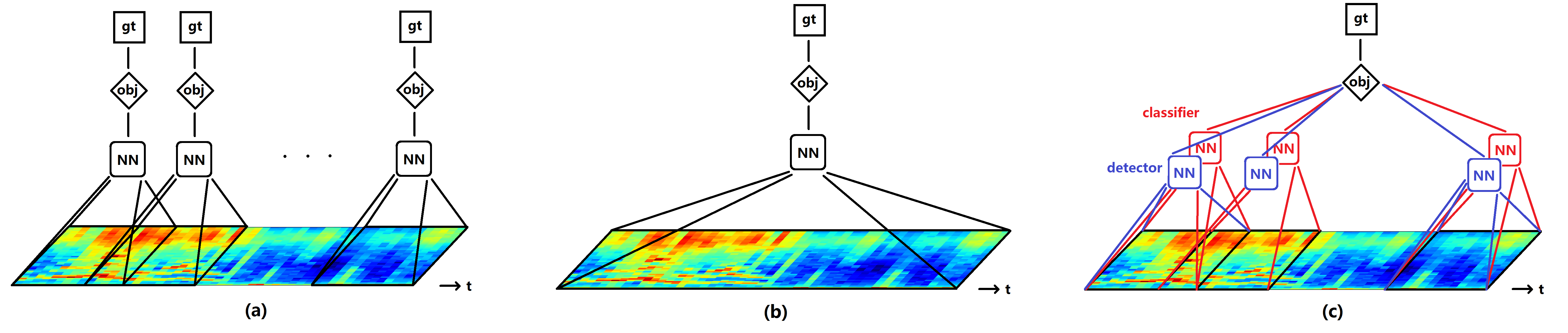}}
  \caption{(a) The BOB model. (b) The global model. (c) The JDC model}
  \label{fig:results}
\end{figure*}

However, the ``tags occur in all blocks'' assumption in the BOB model can not always be satisfied in practice, for example the ``gun shot'' will not occur all the time in an audio clip. The JDC model is able to overcome this problem. Similar to the BOB model, the audio clip is divided into overlapped blocks. The JDC model consists of a detector and a classifier acting on each tag and each block with output value between 0 and 1. The output of the classifier $ y_{km} $ on the $ k $-th tag and the $ m $-th block indicates how possible the $ m $-th block has tag $ k $. The output of the detector $ w_{km} $ on the $ k $-th tag and the $ m $-th block indicates how informative the $ m $-th block is when classifying the $ k $-th tag. If $ w_{km} $ is close to 1 it means the $ m $-th block is informative and should be attended when classifying the $ k $-th tag. If $ w_{km} $ is close to 0 it means the $ m $-th block is uninformative and should be ignored when classifying the $ k $-th tag. 

Compared with the BOB model in Figure 2(a), the JDC model in Figure 2(b) applies a merging operation to calculate the occurrence probability of the $ k $-th tag $ p_{k} $ directly, without calculating the probability over the blocks, hence avoids the ``tags occur in all blocks'' assumption. It is easy to define $ p_{k} = \sum_{m=1}^{M} w_{km}y_{km} $, but this will cause problem. First, the detector $ w_{km} $ and the classifier $ y_{km} $ are no longer distinguishable because they are symmetric. Second, $ p_{k} $ may exceed 1 which is no longer a probability. One way to solve this problem is to enforce the constraint $ \sum_{m=1}^{M} w_{km} = 1 $, or equally, use normalized $ w_{km} $ over blocks to turn the constrained problem to an unconstrained problem. We define $ \widetilde{w}_{km} $ as the normalized detector

\begin{equation} \label{eqDetector}
\widetilde{w}_{km} = \frac{w_{km}}{ \sum_{m=1}^{M} w_{km} }
\end{equation}

Define $ p_{k} $ as the occurrence probability of the $ k $-th tag in the clip 

\begin{equation} \label{eqMergeVal}
p_{k} = \sum_{m=1}^{M} \widetilde{w}_{km} y_{km}
\end{equation}

Define the loss function of the JDC model as
\begin{equation} \label{eqJdcLoss}
loss = \sum_{k=1}^{K} d(p_{k},t_{k})
\end{equation}

\noindent
where $ d(\cdot,\cdot) $ is the binary cross-entropy error.

In the recognition phase, the occurrence probability of the $ k $-th tag in the clip is calculated by equation (4). 

Figure 2(a-c) demonstrates the BOB model, the global model and the JDC model, respectively. The rounded rectangle, diamond and square represents the NN model, objective function and ground truth respectively. The NN can be any kinds of DNNs including CNNs, RNNs and LSTMs. In the training phase, the error back propagates from loss function to both the detector and the classifier in the JDC model (Figure 2(c)). Both the detector and the classifier can be updated according to the gradient. 

The JDC model has the following features:

\begin{itemize}
\item Weakly labelled data can be fed into the model directly without the event level label. 
\item The detector's attend and ignore mechanism simulates the humans perception \cite{bregman1994auditory}.
\item The detector's attend and ignore mechanism facilitates the recognition of short audio events. 
\item Audio event detector is trained without needing the event level label. 
\end{itemize}

\section{Experiments}
\label{sec:Experiments}

We apply both the BOB model and the JDC model on the CHiME home dataset \cite{foster2015chime}. The development set consists of 1946 4 second chunks divided into 5 folds. Tags include ``child speech'', ``male speech'', ``female speech'', ``TV', ``percussive sounds'', ``broadband noise'' and ``other identifiable sounds'' and ``silence'', denoted by `c', `m', `f', `v', `p', `b', `o' and `S', respectively. Another 816 chunks are kept for evaluation. The dataset is weakly labelled without event level labels. 

\subsection{Features}
We apply fast Fourier transform (FFT) with non overlapping Hamming window with size of 1024 on the 16kHz audio files. Mel filter bank features with 40 bins are extracted using librosa toolbox\footnote{https://github.com/librosa/librosa}. Each block contains 11 frames and the hop size between the neighboring blocks is 1.

\subsection{Models}
To compare the BOB model with the JDC model, we simply use the fully connected (FC) neural network to model both the classifier and the detector. This can be extended to any neural network (Section 3).

For the BOB model, the classifier takes one block as input followed by 3 FC layer with 500 units per hidden layer. ReLU activation function and dropout probability of 0.2 are applied to each hidden layer. Sigmoid function is applied on the classifier's output. For the JDC model, the classifier is the same as the BOB model, the detector takes one block as input followed by a mean pool layer along time axis and a single layer linear transform with sigmoid output. Experiments show that the detctor described by a shallow layer NN is sufficient on the ``CHiME Home" dataset. Adam optimizer is used and the learning rate is tuned to 0.0002. The system is built on \textit{Hat} deep learning framework\footnote{https://github.com/qiuqiangkong/Hat}. The source code for the proposed JDC model is available on github\footnote{https://github.com/qiuqiangkong/audio\_tagging\_jdc}.

\subsection{Results}

We use 5-fold cross validation on development set to select the best model (65 epochs for the BOB model and 110 epochs for the JDC model). Then the best model is applied to the evaluation set. The EERs of the BOB model and the JDC model on development set and evaluation set are shown in Figure 3 and Table 1. It is noticeable that the JDC model performs slightly worse than the BOB model on development set (equal error rate (EER) of 17.3\% against 16.7\%) but performs better on the evaluation set (EER of 16.9\% against 19.0\%). This indicates with more training data, the JDC is more effective and suffers less from overfitting.

\begin{figure}[t]
  \centering
  \centerline{\includegraphics[width=\columnwidth]{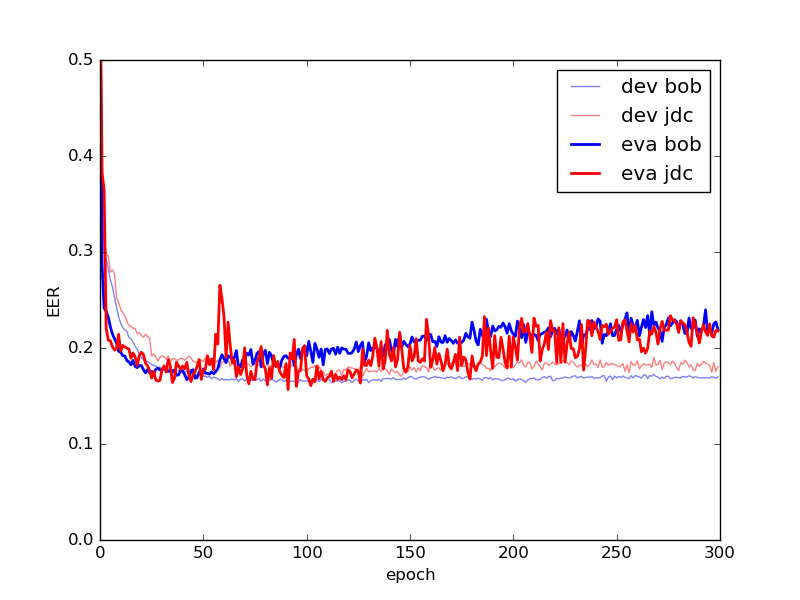}}
  \caption{EER of the BOB model and the JDC model on development set and evaluation set. }
  \label{fig:eer_results}
\end{figure}

\begin{table}[h]
\centering
\caption{EER on development set and evaluation set. }
\begin{tabular}{ |p{2.5cm}|p{2.2cm}|p{2.2cm}| }
 \hline 
 & EER (dev.) & EER (eva.) \\
 \hline
 MFCC + GMM (baseline) & 21.3\% & 20.9\% \\
 \hline
 BOB & \textbf{16.7\%} & 19.0\%  \\
 \hline
 JDC & 17.3\% & \textbf{16.9\%} \\
 \hline
\end{tabular}
\end{table}

\subsection{Visualize Learned Representation}

It is interesting to visualize what the BOB model and the JDC model learn. Figure 4(a) shows the Mel energy spectrogram corresponds to the audio clip in Figure 1. The tags for this clip include children speech, percussion and other sounds, denoted by `c', `p', `o', respectively. Figure 4(b) shows the detector output $ \widetilde{w}_{km} $, where the bright regions indicate informative regions. Figure 4(c) shows the classifier output $ y_{km} $, where the bright regions indicate the probability of the occurrence of the events regardless of the detector. Figure 4(d) is the detection-classification output $ \widetilde{w}_{km} y_{km} $, where bright regions indicates the occurrence probability of the events. Figure 4(e) is the classifier output of the BOB model. Figure 4(f) is the event level label of the clip. Comparing Figure 4(d) and Figure 4(f), the JDC model detects the audio events successfully, even if no event level label data is used in the training phase. However the classifier output of the BOB model in Figure 4(e) does not contain the occurrence times of the events. 

\begin{figure}[t!]
  \centering
  \centerline{\includegraphics[width=\columnwidth]{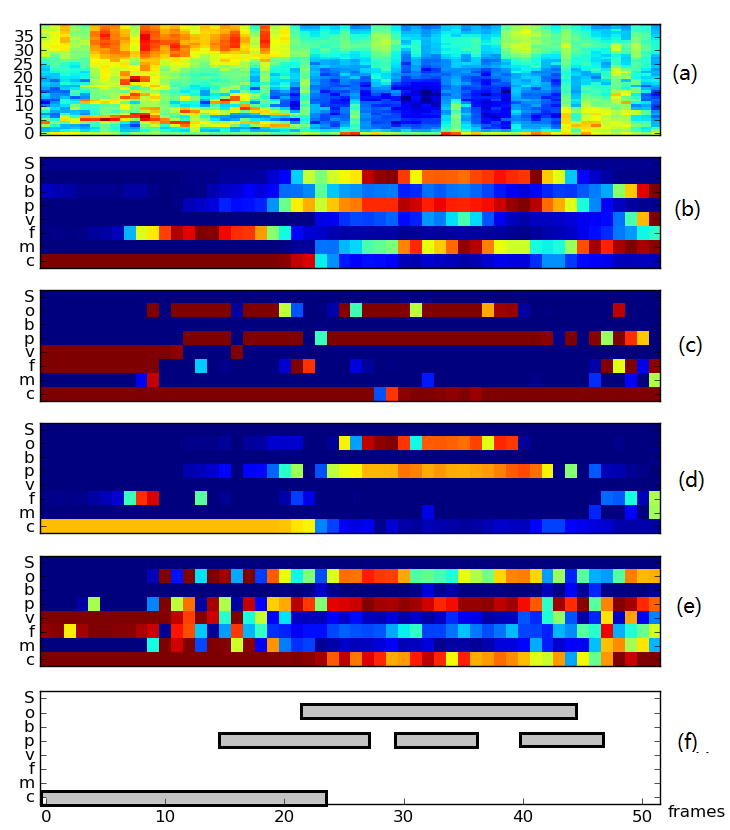}}
  \caption{(a) Mel energy spectrogram of the audio clip in Figure 1. (b) Detector output of the JDC model. (c) Classifier output of the JDC model. (d) Detector-classifier output of the JDC model. (e) Classifier output of the BOB model. (f) Manually labelled event level label. }
  \label{fig:results}
\end{figure}

\section{Conclusion}
\label{sec:Conclusion}

We proposed a \textit{joint detection-classification} (JDC) model to use \textit{weakly labelled data}. By introducing the detector, the model simulates the humans' attend and ignore ability. In addition, the JDC model is able to perform event detection without needing event level label. For the future work, CNNs, RNNs, LSTMs can be used to replace the fully connected NN. In Figure 3 the EER curve of the JDC model oscillates considerably, which may be mitigated by applying different learning rates on the detector and classifier. Expectation maximization (EM) algorithm could be used to replace the joint training method in the JDC model.

%

\bibliographystyle{IEEEbib}
\bibliography{strings,refs}

\end{document}